\documentclass[dvitex,twocolumn,epjc3]{svjour3}

\RequirePackage[T1]{fontenc}

\smartqed  

\RequirePackage{graphicx}
\RequirePackage{mathptmx}      
\RequirePackage{flushend}
\RequirePackage[numbers,sort&compress]{natbib}

\PassOptionsToPackage{dvips}{hyperref}
\RequirePackage[colorlinks,citecolor=blue,urlcolor=blue,linkcolor=blue]{hyperref}

\usepackage{amssymb}
\usepackage{graphicx}
\usepackage{amsmath}
\usepackage{amsfonts}
\usepackage{comment}

\newcommand{\GeV}{\makebox{ GeV}}
\newcommand{\beq}{\begin{equation}}
\newcommand{\enq}{\end{equation}}
\newcommand{\beqa}{\begin{eqnarray}}
\newcommand{\beqast}{\begin{eqnarray*}}
\newcommand{\enqa}{\end{eqnarray}}
\newcommand{\enqast}{\end{eqnarray*}}

\def\GeV{\nobreak\,\mbox{GeV}}

\begin{document}

\title{ Analysis of ATLAS pp elastic measurements  at $\sqrt{s}$=13 TeV  and comparison with TOTEM measurements} 

    \author{ E. Ferreira\thanksref{e1,addr1} \and  T. Kodama\thanksref{e2,addr1,addr2}
    \and  A. K. Kohara\thanksref{e3,addr3}} 
  \thankstext{e1}{e-mail: erasmo@if.ufrj.br}
   \thankstext{e2}{e-mail: kodama.takeshi@gmail.com}         
   \thankstext{e3}{e-mail: anderson.kendi@gmail.com}                              
\institute{Instituto de F\'isica, Universidade Federal do Rio de Janeiro, C.P. 68528,
           Rio de Janeiro 21945-970, RJ, Brazil \label{addr1}
           \and
   Instituto de F\'isica, Universidade Federal Fluminense, Niter\'oi 24210-346, RJ, Brazil \label{addr2}
   \and
   CPHT, CNRS, École Polytechnique, Institut Polytechnique de Paris, 91120 Palaiseau, France 
   \label{addr3}
   }
%
%


                 \maketitle
 

\begin{abstract}
 A comparative description is made of the measurements at LHC of pp elastic scattering at 13 TeV   
by the ATLAS and TOTEM Collaborations. In the total and differential cross sections we show that the 
differences are justified through  single numerical factor. It seems that there is  
no fundamental physical difference, but only a 
difference of normalization between the two experiments.
 We study the  
real and imaginary amplitudes disentangled with the KFK   (Kohara-Ferreira-Kodama) model  and
show  that the properties are similar in qualitative  aspects for both experiments. The real and 
imaginary parts have
different slopes at the origin and present zeros, with distributions that are common to 
several models, with three zeros in the real part and one zero in the imaginary amplitude. 
 A zero in the real part, known as Martin's zero, influences the determination of
the $\rho$ parameter.

\PACS{13.85.Dz \and 13.85.Lg}
\keywords{elastic differential cross-section \and  total cross section \and scattering amplitudes}
 \end{abstract}

  

\section{Introduction\label{Intro}} 

Proton-proton elastic scattering is a fundamental process in studies of hadronic physics.
Due to its nonperturbative  nature the interpretation of data is dependent on models. 
In non-polarized case, $d\sigma/dt$ is described by a complex amplitude with real and imaginary
parts depending on  the variables $s$ and $t$ only, allowing in principle a direct analytical
description .

With respect  to the scale from low to high energies, the   data at TeV energies are  very scarce,
  and there is no  hope  of new experiments in near future. The existing experimental 
information is extremely valuable, and detailed and objective analysis is important   for
 coherent description and interpretation  of what exists. All  effort is  relevant.

The observed elastic differential  cross section is a sum of squares of unknown  independent 
real and imaginary amplitudes, and an effort  towards a possible  dynamical understanding 
of elastic scattering requires that the two parts of the amplitude 
   be    disentangled     and presented explicitly, obviously in a model dependent way.
             

Measurements at $\sqrt{s}=13$~TeV were performed at the LHC by the TOTEM
Collaborations \cite{TOTEM_13_1,TOTEM_13_2,TOTEM_13_3} and ATLAS
Collaboration \cite{ATLAS_13}, yielding significantly different values
of the total cross section.
The TOTEM measurements give $\sigma_{\rm tot}\simeq 110.6$~mb, while
the ATLAS analysis reports $\sigma_{\rm tot}=104.7$~mb.
In our analysis, the amplitudes fitted to the TOTEM data lead, via the
optical theorem, to a total cross section $\sigma_{\rm tot}=111.56$~mb,
consistent with the TOTEM measurements.
As a consequence, the optical theorem implies a corresponding ratio
between the imaginary parts of the forward scattering amplitudes,
\begin{equation}
  \frac{T^{\rm ATLAS}_I(s,t=0)}{T^{\rm TOTEM}_I(s,t=0)} = 0.9385 .
  \label{sigma ratio}
\end{equation}
Starting from this information, we show how the measured differential
cross sections in the two experiments can be consistently compared.

TOTEM Collaboration in LHC has produced two sets of data on elastic pp scattering at 
$\sqrt{s}$=13 TeV in  separate publications   \cite{TOTEM_13_1,TOTEM_13_2,TOTEM_13_3} ,    
covering the following $|t|$  ranges 
\begin{itemize}
  \item Set I -  $|t|= [0.000879 - 0.201041]~ \GeV^2 $ , with N=138 points \cite{TOTEM_13_1} ;
\item  Set II -  $|t|= [0.0384 - 3.82873] ~  \GeV^2 $ , with N=290 points \cite{TOTEM_13_2}  
\end{itemize} 
with a superposition. 
With respect to systematic errors, the two sets of measurement are given with very different 
features: relative systematic errors of about 5$\%$ for I and less than  1$\%$  (except for 
the first 11  points) for Set II.  
   ATLAS Collaboration presents data  \cite{ATLAS_13} with 79 points  in the range
  $|t|=[0.00029-0.4376]~\GeV^2 $,    with relative statistical  errors about 0.2 $ \% $ in the 
 center of   the  measured range and 
relative systematic errors  about  2.2 $ \% $   in the center of the data range. In both
TOTEM and ATLAS  cases 
the errors are   larger in the two extremes of the data.

The purpose of the present work is not to re-evaluate the experimental analyses,
but to provide a comparative and model-based interpretation of the ATLAS and
TOTEM measurements within a common theoretical framework.                                       
\section{\label{Data and Amplitudes} Description of ATLAS data and amplitudes with KFK model}  

\begin{figure*}[b]
     \includegraphics[width=8cm]{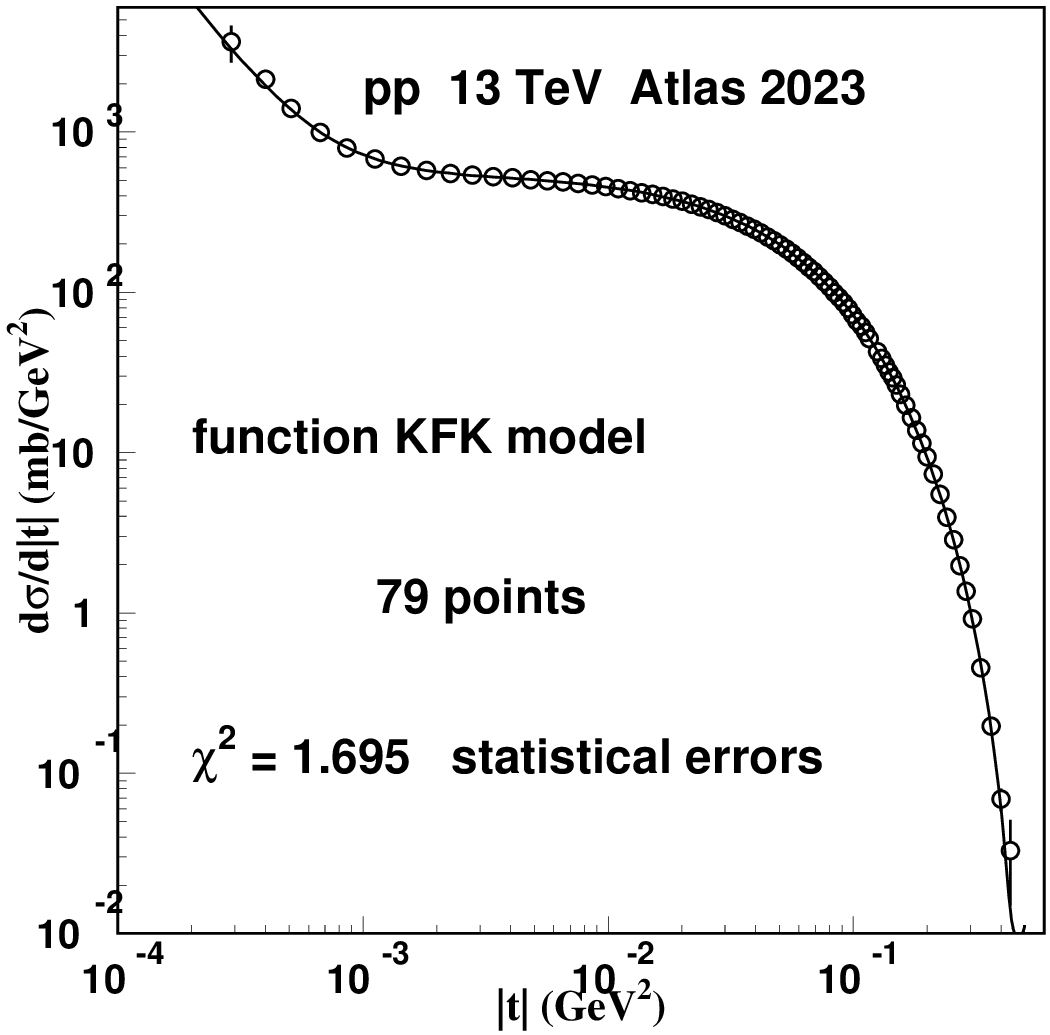} 
\caption {Representation of the 79 points of ATLAS measurements of differential cross section
of pp  elastic scattering at  13 TeV  with a function of the KFK model. Fit and plot with statistical 
errors only.
\label{ATLAS_13_KFK}          }  
\end{figure*}

  TOTEM  data was  studied in detail in the KFK (Kohara-Ferreira-Kodama) model \cite{TOTEM_13_KFK},
with identification of the real and imaginary  amplitudes in the full range of the experiment 
and analysis of b-space (impact-parameter space)  properties. In the present paper we study ATLAS 
 data in similar framework, unfortunately without study of b-space representation because of the small |t|
range.  Fig.(\ref{ATLAS_13_KFK})   shows the whole ATLAS  data
of the differential cross section together  with the corresponding curve described by the KFK model.
  Fit and $\chi^2$ are given with statistical errors only. We note a very good representation.  

\begin{figure*}[b]
     \includegraphics[width=8cm]{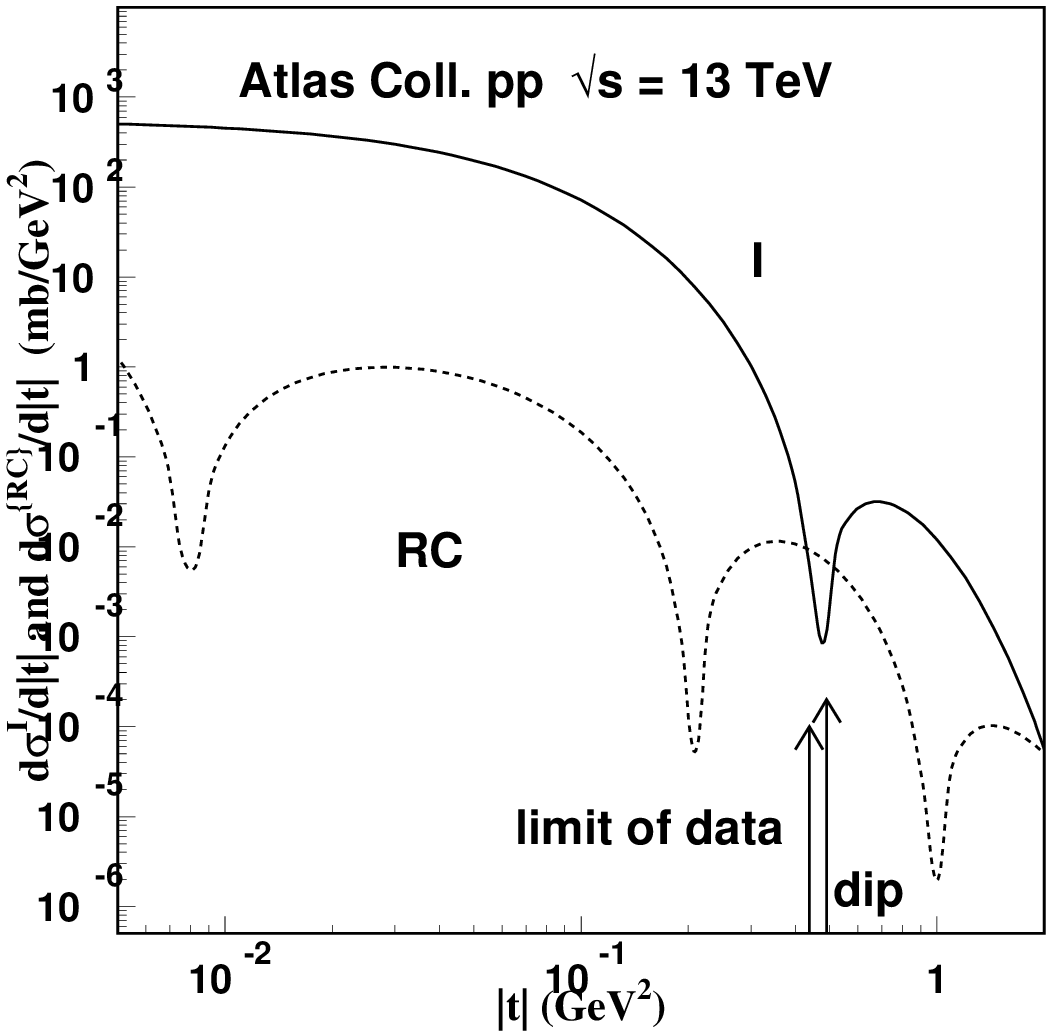}     
     \includegraphics[width=8cm]{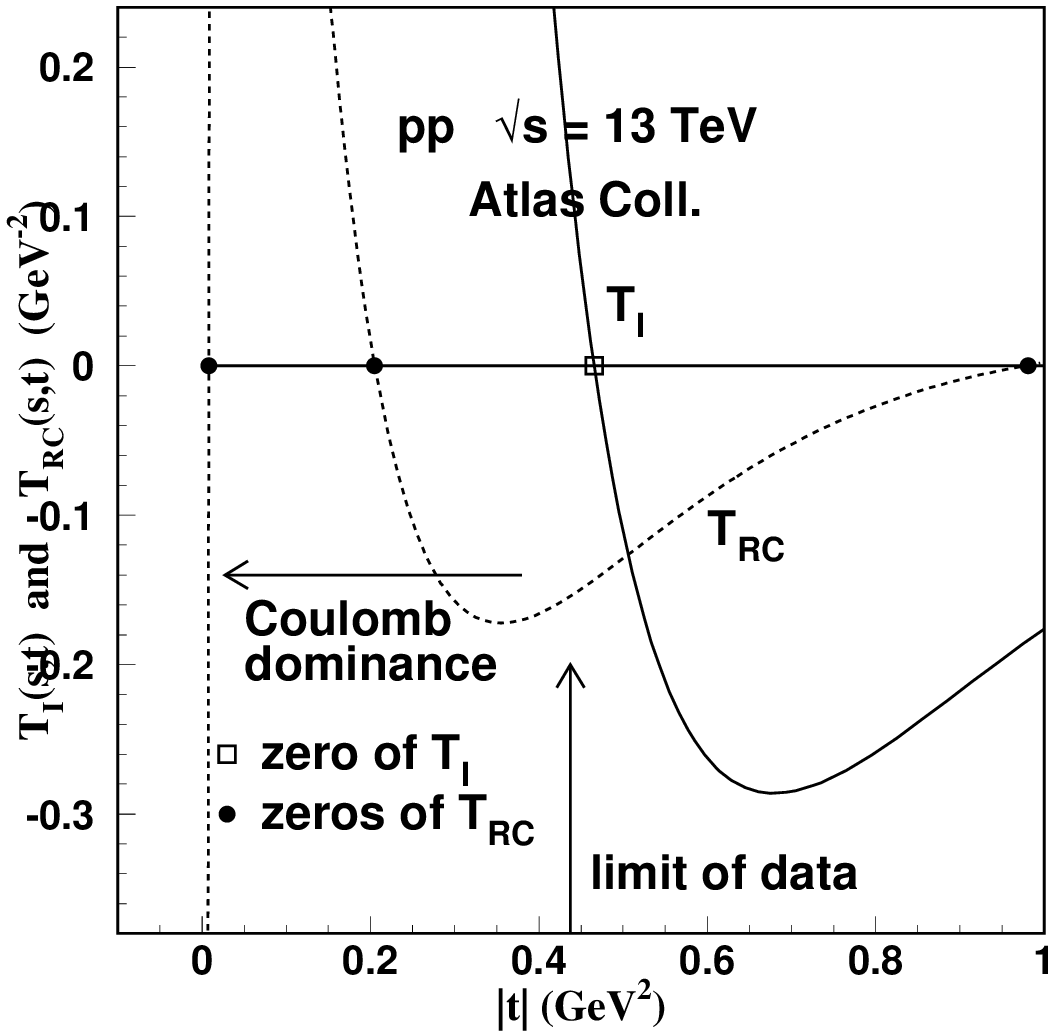}
   \includegraphics[width=8cm]{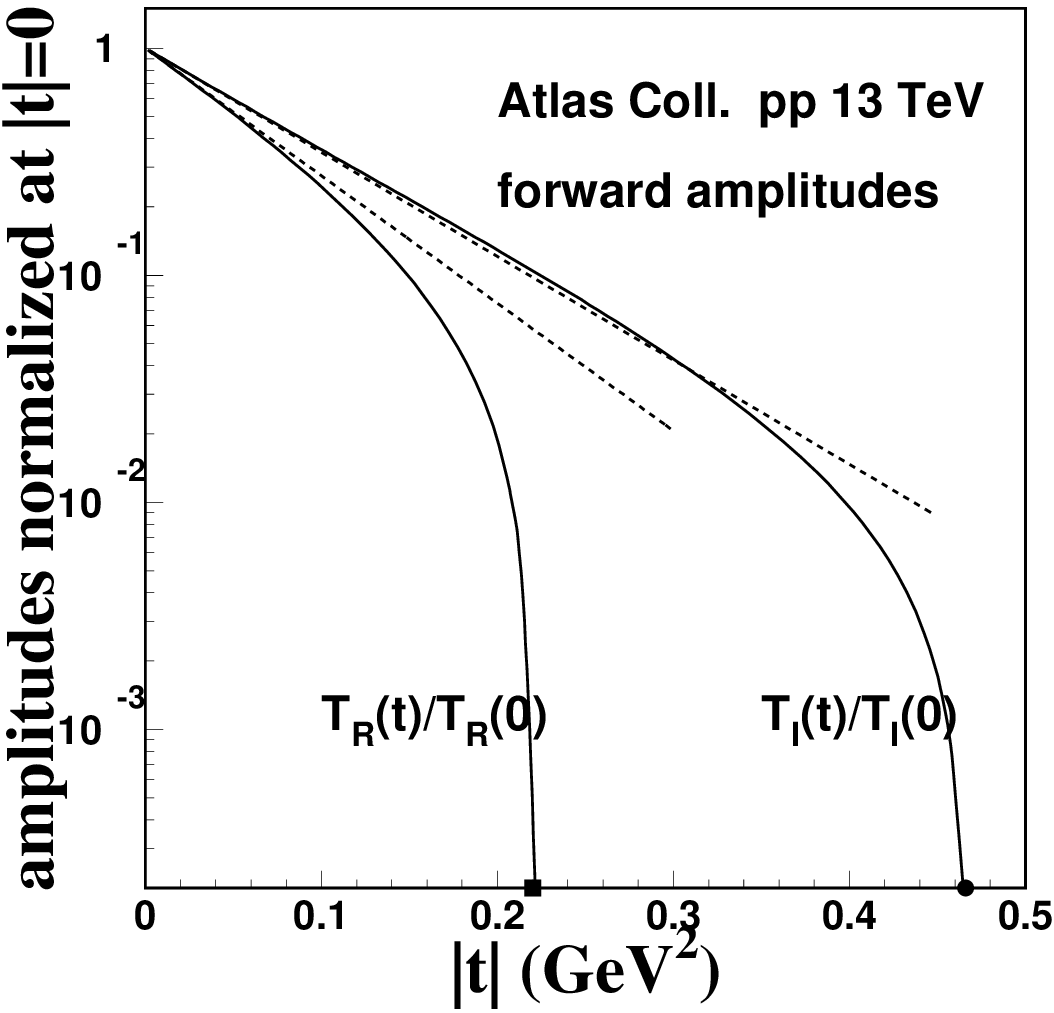}
  \caption {Separation of the imaginary and real amplitudes of $d\sigma/dt$ of ATLAS data 
in the analysis  with the KFK model. The symbol RC means superposition  of real nuclear and Coulomb
amplitudes. In the  first plot we show $d\sigma^{RC}/dt$ and $d\sigma^I/dt$, dips indicating their zeros. 
In the second plot we show the amplitudes.
In the real part the  first zero is in the cancellation with the negative Coulomb amplitude  at $|t|=0.008 \GeV^2$ .
The second real zero is  Martin's zero at $|t|=0.205\GeV^2$. The third real zero
is outside the  data range at  $|t|=0.98\GeV^2$.
The imaginary zero is predicted for $|t|=0.466\GeV^2$, just after the data. The dip in the differential 
cross section is predicted for $|t|=0.47\GeV^2$, also after the end of the data.
In the third plot we show the imaginary and real nuclear amplitudes in the range of very small |t|, with their 
zeros, compared with the behaviour of pure exponentials, with different slopes. 
 \label{amplitudes_ATLAS_13_TeV}  }    
\end{figure*}

Fig.(\ref{amplitudes_ATLAS_13_TeV})  shows the partial  differential cross
sections  (imaginary $d\sigma^I/dt $ and real $d\sigma^R/dt$ parts ), as disentangled with KFK model, 
and the corresponding amplitudes. The predicted  
structure of zeros is similar to the one observed  at  this and lower energies by different models, with
three zeros in the real part and one zero in the imaginary amplitude. The first real zero   $Z_{RC}^{(0)} $ is located in the cancellation 
with the  negative  Coulomb  amplitude at very small |t|.   In table form we show that in  ATLAS data it occurs at |t| = 0.008 $ \GeV^2$,
while  in TOTEM at 13 TeV it is calculated    by KFK at  |t| = 0.0057 $\GeV^2$. The 
second real zero   $Z_{RC}^{(1)} $ is the so-called Martin's zero \cite{Martin}, that in ATLAS  occurs at |t|=0.205 $\GeV^2$, 
as well  observed in the figure, and in the  analysis of TOTEM data at 13 TeV it  appears at |t| = 0.191 $\GeV^2 $. 
This second real zero is crucial for the determination of  the  $\rho$  parameter.
For the ATLAS data the third real zero is predicted by KFK parametrization for   |t| = 0.98 $\GeV^2$ , much 
beyond the data; in TOTEM
data at 13 TeV it is as located at |t|=1.186 $\GeV^2$  \cite{TOTEM_13_KFK} within the data range.
In ATLAS case the single imaginary zero is predicted for |t|=0.466 $\GeV^2$, just after the data ends, 
at |t| =$ 0.4376 \GeV^2$. 
The dip for ATLAS  position is  predicted  by KFK at |t| = 0.470$\GeV^2$, close to the predicted imaginary zero, 
as occurs for  all 
energies, at and above  ISR.  Fig.(\ref{ATLAS_13_KFK}) shows that actually  the ATLAS  data seems to tend to   a 
dip near and after its limit.             

\begin{figure*}[b]
      \includegraphics[width=8cm]{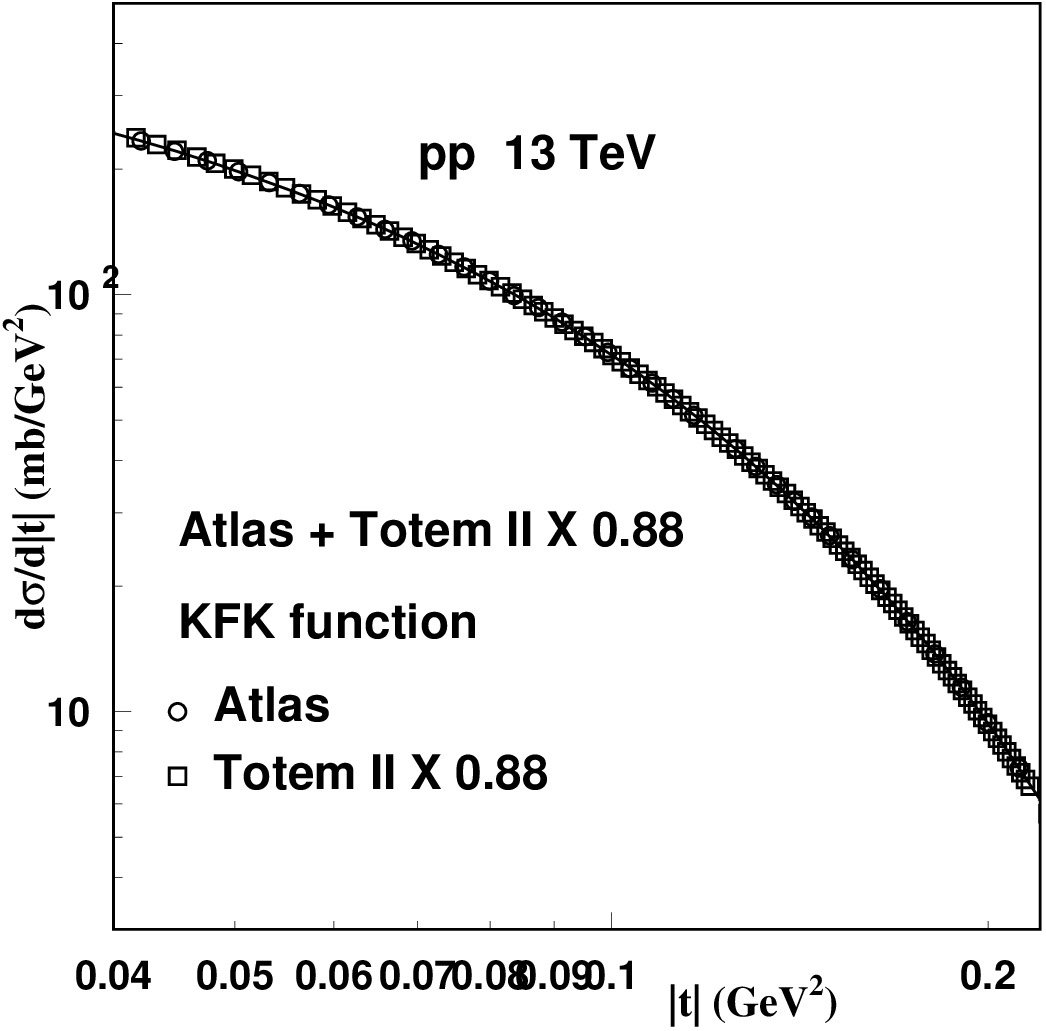}    
     \includegraphics[width=8cm]{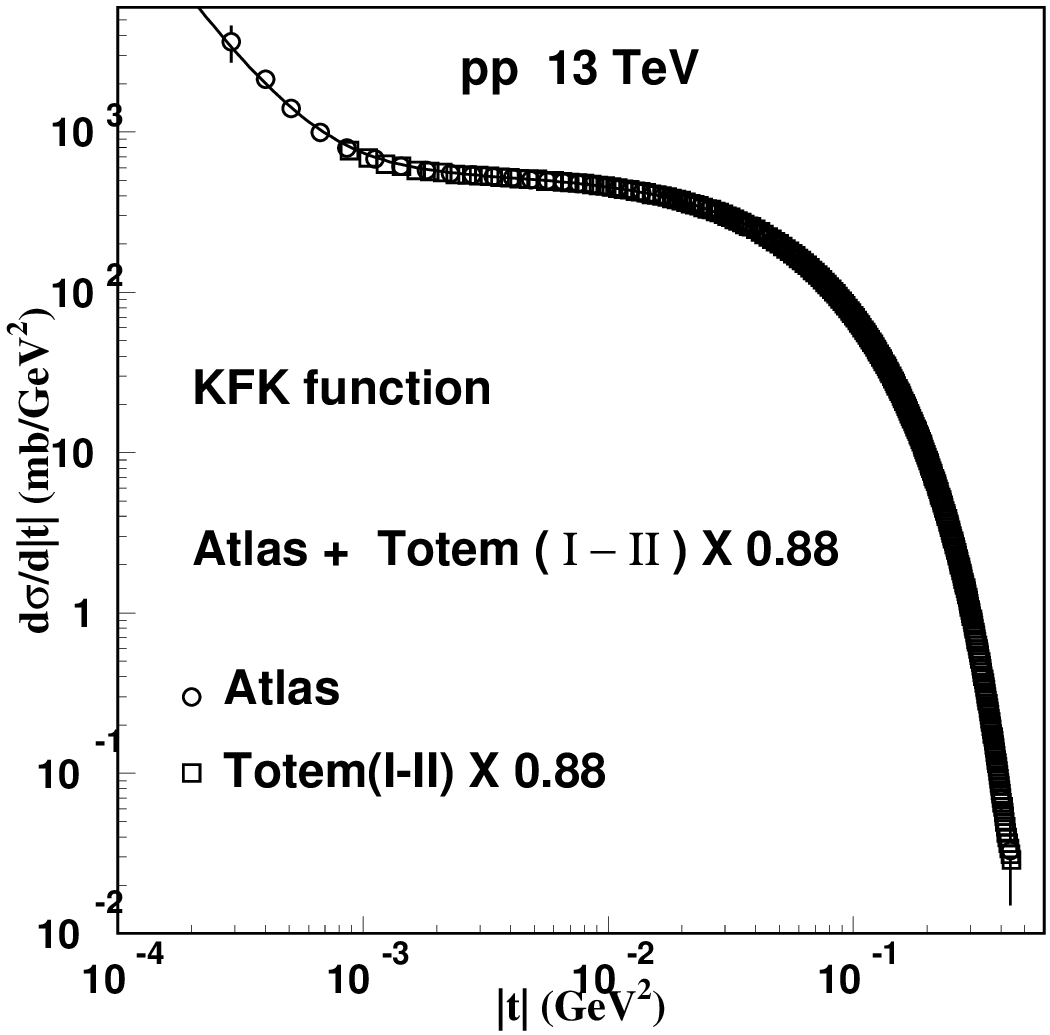} 
\caption {ATLAS  and TOTEM  data   differ by a factor 
0.88 in $d\sigma/t$
that is  determined by the square of the ratio of the total cross sections 
$\sigma^{\rm ATLAS}/\sigma^{\rm TOTEM}=104.7/111.56=0.9385$ that is the ratio of the t=0    
imaginary amplitudes.    
         In the first plot a range is chosen to show the  precision
  of the factor 0.88. The second plot includes all 79 points
of ATLAS data together with proper junction of  TOTEM data of Sets I and II. 
The fitting  function of   data ATLAS  is given by  KFK model,
the same used in Figs. (\ref{ATLAS_13_KFK},\ref{amplitudes_ATLAS_13_TeV}).  
  \label{adjusting factor}      }     
 \end{figure*}

   To obtain  a relation between the differential cross sections,
we first  compare the values of the total  cross sections  that are 104.68 mb for ATLAS and 111.56 mb 
for TOTEM. The ratio is 0.938.   
By the optical theorem this is the ratio of the |t| = 0 
 imaginary amplitudes as shown in Eq.(\ref{sigma ratio}).
 As the differential cross sections  are  given by combination of squared amplitudes,   
we may expect a square of this factor, namely $(0.938)^2\approx 0.88 $,  optimistically   
in the whole t range. Thus   
\begin{equation}
 \frac{  d\sigma^{\rm ATLAS}/dt } { d\sigma^{\rm TOTEM}/dt } ~~~~  \approx 0.88  ~~~~~, ~ ~~ all ~~ |t|  ~~.       
\label{dsigdt ratio}
\end{equation}
 This hypothesis of a simple normalization factor between the two
experiments  is tested  in Fig. (\ref{adjusting factor}). This scaling is not meant as a statement of exact factorization at all |t|,
but as an effective empirical relation within the measured ranges.
 In the left-hand side  we choose a range |t| = (0.04-0.22) $\GeV^2$ with Set-II of TOTEM 
where both experiments  have good quality  and plot them together, with the 
  multiplying factor 0.88.  Visually a very nice superposition in this range is obtained. 
In the  right hand side plot the same factor 0.88 is applied for the whole  ATLAS  data,
using a proper junction  of Sets  I (56 points)  and II (214 points) of TOTEM  data. 
Again a very good comparison is observed, particularly if we do not consider with rigor
 some  points in the extremes that have large error bars, particularly  in ATLAS experiment. 

We may confirm the  over-all factor 0.88 for $d\sigma/dt$ examining the  integrated   elastic   cross section  $\sigma_{\rm el} ^{ATLAS} $. 
The ATLAS article informs the value of the ratio $\sigma_{\rm el}^{ATLAS}/\sigma_{\rm tot}=0.257\pm0.008\pm0.009$.
With the value $\sigma_{\rm tot}=104.68\pm1.08\pm0.12$ mb  we calculate the value $\sigma_{\rm el}^{ATLAS}=26.903\pm0.831$ mb.  
 The KFK model calculation  for the same ATLAS data gives very similar value $\sigma_{\rm el}^{ATLAS}=27.2416$ mb.
On the other hand the value with TOTEM data is   $\sigma_{\rm el}^{TOTEM}= 31.0972$ mb. Thus the ratio between
the two LHC experiments is
\begin{equation} 
\frac{\sigma_{\rm el}^{\rm ATLAS}}{\sigma_{\rm el}^{\rm TOTEM}}= 0.865  ~~~{\rm{or}} ~~~0.876  ~
\label{ratio_integrated}                                              
\end{equation}
 As we like to expect, these two numbers are very close to the square  of the ratio  of the two cross sections , namely   
$0.938^2=0.88$, as written in Eq.(\ref{dsigdt ratio}) 
and used in Fig.(\ref{adjusting factor}).
Naturally we can be satisfied with the conclusion that ATLAS and TOTEM measurements are compatible, up to 
a normalization factor, in the directly observed ranges  of  differential cross sections. 
However the quantity $\rho$ is dependent on  the model used for the extraction of its value, that 
actually  receives different definitions. 
 Thus, for example,  the ATLAS experimental paper  
\cite{ATLAS_13} gives in the abstract the value $\rho=0.098\pm0.011$, while the KFK model 
calculates    $\rho=0.091\pm0.004$  for the same data. We  do not see logical 
relation between these numbers and the ratios of the total or  differential cross sections. Since the $\rho$ parameter relies on model assumptions for the real and imaginary parts of the scattering amplitude, it is safer to regard this quantity as a model-dependent parameter rather than as a direct physical observable. This situation contrasts with the total cross section, which is determined through the optical theorem and is therefore directly linked to the imaginary part of the amplitude at the origin. Moreover, in the forward region the imaginary amplitude is typically an order of magnitude larger than the real one, making the extraction of $\rho$ particularly sensitive to modeling assumptions.
We understand that the two calculations applies different definitions for the quantitity.
Similarly with the slope, that in KFK refers to  two different amplitudes, as shown in the appendix. 

We wish to emphasize that a descriptive model should disentangle imaginary and real amplitudes,
otherwise it cannot  define and evaluate properly the $\rho$ value. This separation 
 is not made in   the case of  the basic Eqs. of the ATLAS experimental paper where the real and 
imaginary  amplitudes are not independent, forced to follow the same slopes, and do not
allow a necessary  Martin's    zero in the real nuclear amplitude.

Important  analysis can be made comparing the real and imaginary amplitudes in the 
two experiments. In Fig.(\ref{amplitudes_ATLAS_13_TeV}) we present the amplitudes 
and partial cross sections of ATLAS measurements calculated with the KFK model. 
The configuration of the description is similar to TOTEM data  (see Fig.6 of Ref.5), 
with the limitation
in the data range. We must remark that in KFK the imaginary and real amplitudes have
independent and different slopes, while in ATLAS article the amplitudes run parallel. Besides,
KFK has a zero in the real amplitude (the Martin's Zero) in the measured data range.   

  In  {\ref{KFK Model}  we give basic formulas of the KFK Model  and applications to the  
  comparison of data of ATLAS and TOTEM Collaborations.

\section{\label{Final Remark} Final remarks}

In this work we have presented a comparative analysis of the ATLAS and TOTEM
measurements of elastic pp scattering at $\sqrt{s}=13$~TeV within the
framework of the KFK model. The observed differences between the two
experiments in the measured differential cross sections can be consistently
accounted for by a single global normalization factor, directly related to
the ratio of the total cross sections through the optical theorem.

The ATLAS data are limited in the accessible $|t|$ range and therefore do not
reach the dip--bump region observed in the TOTEM measurements, nor do they
allow a detailed investigation of Coulomb--nuclear interference effects at
very small momentum transfer. Nevertheless, within the measured range, the
qualitative behavior of the real and imaginary parts of the scattering
amplitude extracted from ATLAS data is similar to that obtained from the
TOTEM analysis.

Within the KFK framework, the separation of real and imaginary amplitudes,
each characterized by independent slopes and zeros, plays a crucial role in
the interpretation of the data. In particular, the presence of a zero in the
real part of the nuclear amplitude (Martin’s zero) has a direct impact on the
determination of the $\rho$ parameter. In this context, the ATLAS data are
consistently described by the value
\begin{equation}
{\rm \rho}^{\rm ATLAS} = 0.091 \pm 0.004 \, ,
\end{equation}
which illustrates the model-dependent nature of this quantity and the
importance of the underlying assumptions used in its extraction.

The Appendix summarizes the parameters, derived quantities, and integrated
cross sections obtained within the KFK model for both ATLAS and TOTEM data,
highlighting their qualitative similarity and supporting the conclusion that
the two measurements are compatible within a coherent phenomenological
description.

\section{Acknowledgements}
E.F.  wishes to thank the Brazilian agency CNPq   for financial support.   
 Part of the present work was developed under the project INCT-FNA Proc. No. 464898/2014-5.

\appendix    

\section{\label{KFK Model} Basic formulas of the KFK model used in the comparative   
analysis  of ATLAS and TOTEM data at 13 TeV}

The amplitudes in the KFK model \cite{TOTEM_13_KFK} are phenomenologically based 
on the idea of the Stochastic Vector Model  \cite{Dosch},\cite{Dosch-Simonov},
\cite{SVM}. This phenomenology is   basically
constructed through   b-space profile functions, that give insight for 
geometric aspects of the collision, playing role in the eikonal representation, 
where unitarity constraints have  interesting formulation. 
The structure of the real and imaginary amplitudes of the KFK model are 
explained in several papers, such as in the  description of the Totem data
at  13 TeV \cite{TOTEM_13_KFK}.

Important property of KFK model is that the imaginary and real amplitudes    
can be analytically Fourier-transformed between  b- and  t- spaces. 
With   explicit form   of the amplitudes  in (s,t) space, the experimental 
data are fitted to obtain values of parameters.

The nuclear   amplitudes  are    written 
   $T_R^N(s,t)$ and $T_I^N(s,t)$ with  forms containing perturbative and
non-perturbative terms  
\begin{eqnarray}
\label{hadronic_complete}
 T_{K}^{N}(s,t) =\alpha_{K}(s)\mathrm{e}^{-\beta_{K}(s)|t|}+ \lambda _{K}(s)\psi    
_{K}(\gamma _{K}(s),t)  
\end{eqnarray}
with K = R,I indicating either the real or the imaginary part of the complex amplitude.

The perturbative parts have two parameters, $\alpha_{K}(s)$ and $\beta_{K}(s)$, in each amplitude;
the nonperturbative parts have two parameters, $\lambda_{K}(s)$ and $\gamma_{K}(s)$,  in each
amplitude.

   The non-perturbative shape function is written for each K=R,I 
 \begin{eqnarray}
 \label{psi_st}
&&\psi _{K}(\gamma _{K}(s),t) \\
&=&2~\mathrm{e}^{\gamma _{K}}~\bigg[{\frac{\mathrm{e}^{-\gamma _{K}\sqrt{%
1+a^2|t|}}}{\sqrt{1+a^2 |t|}}}-\mathrm{e}^{\gamma _{K}}~{\frac{e^{-\gamma
_{K}\sqrt{4+a^2|t|}}}{\sqrt{4+a^2 |t|}}}\bigg]~,   \nonumber
\end{eqnarray}%
%
with the property 
\begin{equation}
\psi _{K}(\gamma _{K}(s),t=0)=1~   .  \label{psinorm2}
\end{equation}%

The quantity $a$, called correlation length, represents properties 
of the QCD vacuum, where it sets the scale for the loop-loop correlation,
with determination in static (Euclidean space) lattice calculation \cite{DiGiacomo} as 
0.25-0.30 fm.

The  four parameters of  each of the I,R amplitudes  give  a regular 
proposal for the separation, and with very good description for $d\sigma/dt$
data for all t. The example of the TOTEM experiment at 13 TeV \cite{TOTEM_13_KFK} is 
recalled below.

  The complete  amplitudes  
contain the nuclear and the Coulomb parts as 
\begin{equation}
T_{R}(s,t)=T_{R}^{N}(s,t)+\sqrt{\pi }F^{C}(t)\cos (\alpha \Phi )~,
\label{real}
\end{equation}%
and 
\begin{equation}
T_{I}(s,t)=T_{I}^{N}(s,t)+\sqrt{\pi }F^{C}(t)\sin (\alpha \Phi )~,
\label{imag}
\end{equation}%
where $\alpha ~$is the fine-structure constant, $\Phi (s,t)$ is the 
interference  phase (CNI)  and $F^{C}(t)$ is related with the proton form factor 
\begin{equation}
F^{C}(t)~=(-/+)~\frac{2\alpha }{|t|}~F_{\mathrm{proton}}^{2}(t)~,
\label{coulomb}
\end{equation}%
for the pp$/$p$\mathrm{{\bar{p}}}$ collisions. The proton form factor is
taken as%
\begin{equation}
F_{\mathrm{proton}}(t)=[t_{0}/(t_{0}+|t|)]^{2}~,  \label{ff_proton}
\end{equation}%
where $t_{0}=0.71\ $GeV$^{2}$.  

The CNI phase is implemented  in Cahn's 
 formalism, with proton form factor adapted to different slopes in the R,I amplitudes
\cite{TOTEM_13_KFK}.

 In our normalization the elastic differential cross section is written 
\begin{eqnarray}  \label{Sigma_diff}
\frac{d\sigma(s,t)}{dt}&=& (\hslash c)^2[T_I^2(s,t)+ T_R^2(s,t)] \\
&=& \frac{d\sigma^I(s,t)}{dt} + \frac{d\sigma^R(s,t)}{dt} ~ ,  \notag
\end{eqnarray}
with  $T_{R}(s,t)$ and $T_{I}(s,t)$ in $\GeV^{-2}$ units, and 
$$  (\hslash c)^2~ =  ~ 0.389379 ~{\rm{mb}}\GeV^2 ~ . $$

The parameters obtained in the fit of the 79 points of ATLAS data with statistical errors 
shown  in Fig.(\ref{ATLAS_13_KFK}) are given  in Table {\ref{KFK_parameters}.
In the same table we include also the original parameters for the 428 points of the TOTEM
data.
  
{\small  
\begin{table*}[ptb]   
\caption  {Parameters of the amplitudes in the KFK model  for energy 13 TeV obtained for ATLAS and TOTEM  experiments.
  The QCD quantity related to  correlation function is 
 $a^2=2.1468\pm 0.0001~ \GeV^{-2} =  (1.4652 ~ \GeV^{-1}\pm 0.0002)^2=(0.2891 \pm 0.0002 ~ {\rm {fm}})^2 $ ,
where $a$ is called correlation length. 
 The quantities  $\gamma_I$ and $\gamma_R$ characteristic of the non-perturbative shape functions in 
Eq.(\ref{psi_st})  are dimensionless, while $\alpha_K$, $\beta_K$ and $\lambda_K$ have units $\GeV^{-2}$.  
The index  $K$ means $I,R$.     
\label{KFK_parameters}  
 }
    \begin{tabular*}{\textwidth}{@{\extracolsep{\fill}}ccc|cccc|cccc@{}}  
     \hline
   \multicolumn{3}{c}{Experiment }&\multicolumn{4}{c} {Imaginary Amplitude}   & \multicolumn{4} {c} {~ Real ~  Amplitude ~   } \\ \cline{1-11} 
 $ \sqrt{s}$ & N pts & |t| range      &$\alpha_I$ &$\beta_I$   &$\lambda_I$  & $\gamma_I$  &  $\alpha_R$   & $\beta_R$    & $\lambda_R$   &$\gamma_R$  \\ 
  TeV        &       &                &$\GeV^{-2}$&$\GeV^{-2}$ & $\GeV^{-2}$ &             &  $\GeV^{-2}$  & $\GeV^{-2}$  & $\GeV^{-2}$   &             \\ 
\hline 
  13 TeV      &  79   & 0.00029-0.4376 &$15.073 $  &$4.4025 $  &$22.846 $     &$7.9021$     &   $0.3022 $   & $1.6075 $    &$3.1478 $      & $7.5409$    \\
ATLAS       &       & $\GeV^2$       &$\pm0.0036$&$\pm0.0023$&$\pm0.0065$   &$\pm0.0027$  &   $\pm0.0598$ & $\pm 0.7705$ &$\pm0.00871$   & $\pm0.3995$   \\ 
  \hline
   13 TeV      & 428  &0.000879-3.82873&$15.701 $  &$4.323 $  &$24.709 $     &$7.819 $     &   $0.2922 $   & $1.540 $     &$4.472 $       & $7.503 $    \\
  TOTEM       &       & $\GeV^2$       &$\pm0.001$ &$\pm0.001$ &$\pm0.002$    &$\pm0.0005 $ &   $\pm0.0005$ & $\pm 0.003$  &$\pm0.003$     & $\pm0.006$  \\ 
\hline
\end{tabular*}
\end{table*}  
}     
    
  
{\small
\begin{table*}[ptb]
\caption{Quantities derived from the solution of the fitting of
  the  13 TeV data, of the ATLAS and TOTEM experiments, with the parameters  given in Table \ref{KFK_parameters}. 
The quantities $Z_I$,  $Z_{RC}^{(0)}$,   $Z_{RC}^{(1)}$  and $Z_{RC}^{(2)}$  are the locations ($|t|$ values) of the zeros 
of the imaginary and real amplitudes (added to Coulomb). Properties of the amplitudes are presented in the Appendix . }
   \label{quantities_derived}              
    \begin{tabular}{@{\extracolsep{\fill}}cccccccccccc@{}}
      \hline  
 \multicolumn{1}{c}{Experiment} & \multicolumn{3}{c} {Imaginary Amplitude} & \multicolumn{5}{c} {Real Amplitude}    \\ \cline{1-9}
      &  $\sigma$     & $Z_I$    &$B_I$           &$\rho$         & $Z_{RC}^{(0)}$ & $Z_{RC}^{(1)}$   & $Z_{RC}^{(2)}$  &$B_R$  \\ 
      &   mb          &$\GeV^{2}$&$\GeV^{-2}$     &$      $       & $\GeV^{2}$     & $\GeV^{2}$    & $\GeV^{2}$    &$\GeV^{-2}$  \\  \hline 
ATLAS &$104.68\pm0.03$&$0.466$   &$21.105\pm0.005$&$0.091\pm0.004$&$0.0080$&$0.205$ &$0.981$ &$25.87\pm1.17$ \\ \hline 
TOTEM &$111.56\pm0.01$&$0.460$   &$21.052\pm0.002$&$0.118\pm0.001$&$0.0057$&$0.191$ &$1.186$ &$26.39\pm0.02$ \\ \hline 
\end{tabular}  
\end{table*}        }


{\small
\begin{table*}[ptb]
   \caption{ Integrated quantities from squared  amplitudes calculated with KFK for the 13 TeV data of ATLAS and TOTEM   experiments. 
    \label{integrated_quantities}  }    
    \begin{tabular}{@{\extracolsep{\fill}}ccccccccc@{}      }         \hline
\multicolumn{1}{c}{Energy}&\multicolumn{1}{c}{Total}&\multicolumn{1}{c}{Elast.}&\multicolumn{1}{c}{Inel.}&\multicolumn{2}{c}{Dip(RC)}&\multicolumn{2}{                                                                       c}{Bump(RC)}&\multicolumn{1}{c}{Ratio}\\ \cline{1-9}
$\sqrt{s}$&$\sigma$&$\sigma_{\rm el}$&$\sigma_{\rm inel}$&$|t|_{\rm dip}$&${\rm h_{dip}}$&$|t|_{\rm bump}$&${\rm h_{bump}}$&${\rm h_{bump}/h_{dip}}$\\  
   TeV    &     mb &    mb           &  mb               & $\GeV^{2}$    & mb/$\GeV^2$     &  $\GeV^{2}$      & mb/$\GeV^2$ &  \\ \hline    
ATLAS    & 104.68 &   27.24         & 77.44             & $ 0.47$       & $0.008   $  &  $ 0.668$  &  $ 0.033 $  & $ 4.13 $   \\ \hline 
 TOTEM    & 111.56 &   31.10         & 80.46             & $ 0.47$       & $0.027    $  &  $ 0.640 $  &  $ 0.047 $  & $ 1.77 $   \\ \hline 
\end{tabular}  
 \end{table*}      }      

Fig.(\ref{full_t})  shows  the   representation in the KFK model of the differential cross
sections of TOTEM and ATLAS data, extended to the full |t| range.  In the ATLAS
 case the data are limited in range, and shown represented as a fictitious  extension
in dashed line.  As we shown  a similar tail, KFK predicts also here  the
importance of the real amplitude  for very large |t|, that must be studied at all energies.

\begin{figure*}[b]
      \includegraphics[width=8cm]{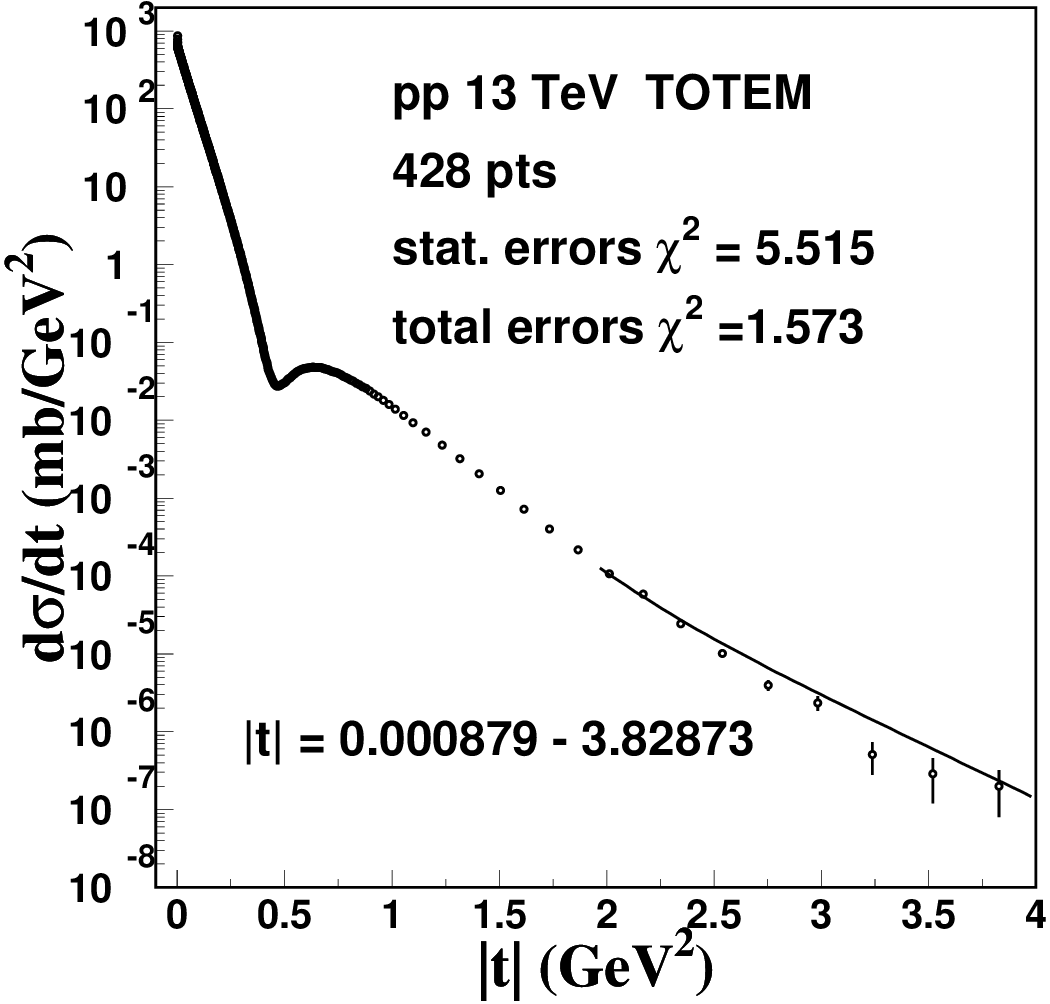} 
\includegraphics[width=8cm]{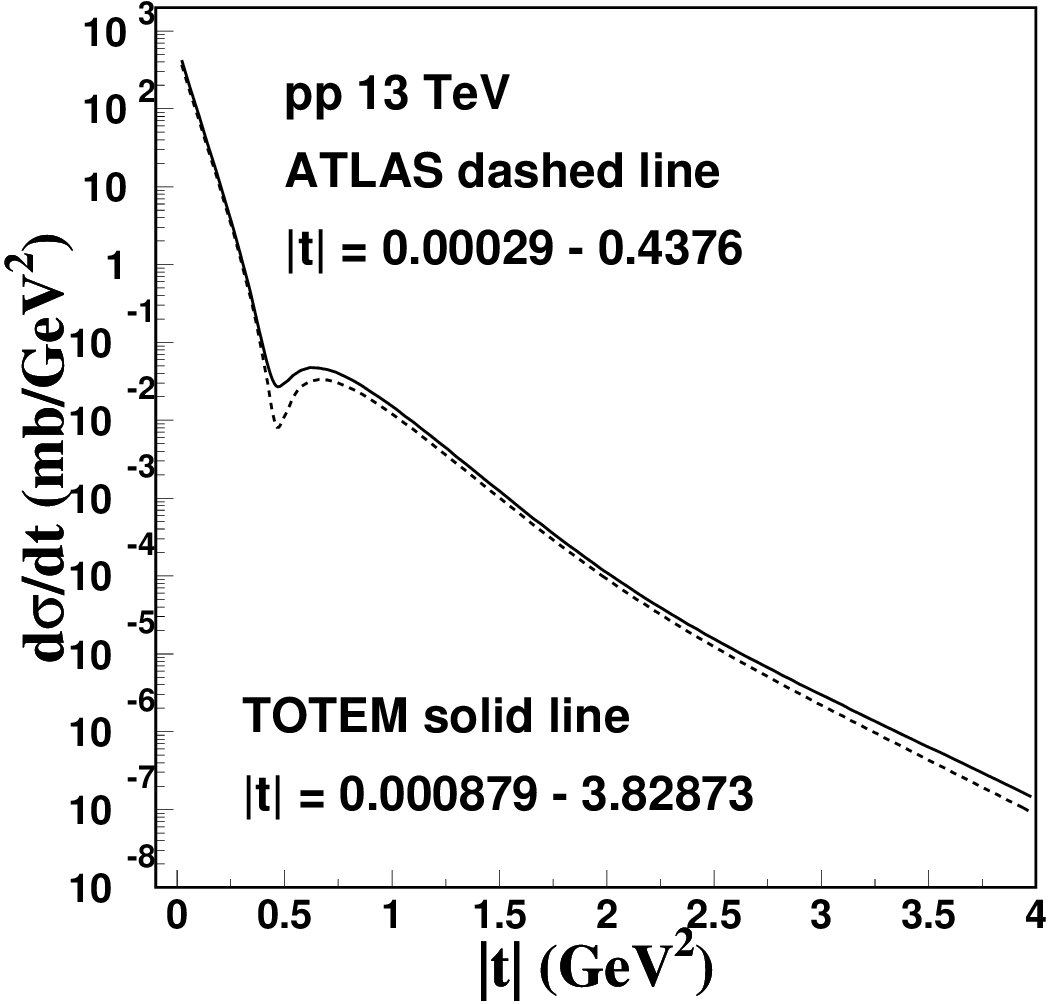} 
\caption {Full ranges of representations of ATLAS and TOTEM data  by KFK Model. 
In the LHS only the TOTEM data and representative curve are shown. In the RHS
the comparative lines are shown for the KFK calculation with the parameters
for ATLAS and TOTEM  data given in Table \ref{KFK_parameters}. The dashed line 
has a  fictitious extension for the  ATLAS case. The dipper $d\sigma /dt$ shape in the 
dashed line is due to the proximity of the values of zeros imaginary $Z_I=0.466 $ and  
real $Z_{RC}^{(2)}=0.981$ in ATLAS data as compared to the corresponding zeros in TOTEM data.
  \label{full_t}   }     
 \end{figure*}


   \end{document}